\documentstyle[aps,epsfig]{revtex}
\title{ Deformation effects in the $^{28}$Si+$^{12}$C and
$^{28}$Si+$^{28}$Si reactions} 

\author{C. Bhattacharya$^{a,e}$, M. Rousseau$^{a}$, C. Beck$^{a}$, V. Rauch$^{a}$, R.M.
Freeman$^{a}$, F. Haas$^{a}$, O. Dorvaux$^{a}$, K. Eddahbi$^{a}$,
P. Papka$^{a}$, O. Stezowski$^{a}$, S. Szilner$^{a}$, D. Mahboub$^{a}$, A. Szanto de Toledo$^{b}$,
A. Hachem$^{c}$, E. Martin$^{c}$, S.J. Sanders$^{d}$ }

\address{ $^a$  Institut de Recherches Subatomiques, F-67037 Strasbourg,
Cedex 2, France
$^b$ Instituto de F\'{i}sica da Universidade de S\~ao Paulo,
S\~ao Paulo, Brazil 
 $^c$ Universit\'e de Nice-Sophia-Antipolis, Nice, France
 $^d$ University of Kansas, Lawrence, KS 66045, USA
 $^e$ Present address : VECC Calcutta, India}
\tighten
\begin{document}

\maketitle
\begin{abstract}
{The possible occurence of highly deformed configurations is investigated
in the $^{40}$Ca and $^{56}$Ni di-nuclear systems as formed in the $^{28}$Si+
$^{12}$C,$^{28}$Si reactions by using the properties of emitted light charged
particles. Inclusive as well as exclusive data of the heavy fragments and their
associated light charged particles have been collected by using the {\sc ICARE}
charged particle multidetector array. The data are analysed by Monte Carlo
CASCADE statistical-model calculations using a consistent set of parameters
with spin-dependent level densities. Significant deformation effects at high
spin are observed as well as an unexpected large $^{8}$Be cluster emission of a
binary nature. } 

\end{abstract}
\vspace{.8cm}

{\bf {keywords:}} {Fusion-fission, Nuclear deformation, Exclusive light charge
particle
measurements}

\section{Introduction}

\vspace{-.4cm}

Extensive efforts have been made in recent years, to understand the decay of
{\bf light} di-nuclear systems (A$_{CN}$ $\leq$ 60) formed through low-energy
(E$_{lab}$ $\leq$ 10 MeV/nucleon) heavy-ion reactions [1].  In most of the
reactions studied, the observed fully energy-damped  yields of the fragments
have been successfully explained in terms of a fusion-fission mechanism [1]
with the noticeable exception of $^{28}$Si+$^{12}$C reaction for which the
deep-inelastic orbiting mechanism has been found to be particularly competitive
[2]. Strong resonance-like structures have been observed in elastic and
inelastic excitation functions of some specific reactions (such as
$^{24}$Mg+$^{24}$Mg or $^{28}$Si+$^{28}$Si) indicating the presence of shell
stabilized, highly deformed configurations in the ($^{48}$Cr and $^{56}$Ni)
compound systems~\cite{zurmuhle83,betts81}. The
investigation of the structure of the doubly-magic $^{56}$Ni nucleus has
recently become a subject of great interest
~\cite{rudolph99a,mizusaki00,svensson99,yu99,macchiavelli00a}. In a recent experiment using the {\sc EUROGAM} multidetector array, the present
collaboration studied the possibility of preferential population of highly
deformed bands in the symmetric fission channel of the $^{56}$Ni compound
nucleus (CN), produced through the $^{28}$Si+$^{28}$Si~\cite{nouicer99,beck00}
reaction at E$_{lab} = 112$ MeV, which corresponds to the energy of the
conjectured J$^{\pi}$ = 38$\hbar$ quasi-molecular resonance~\cite{betts81}. 

\noindent
The study of light charged particle (LCP) emission is a good tool in exploring
nuclear deformations and other properties of hot rotating nuclei at high spins
and angular momenta~\cite{viesti88,govil87,larana88,huizenga89,govil98}. The present work has been performed
with the aim  to investigate the possible
occurence of highly deformed configurations  of the $^{40}$Ca and $^{56}$Ni
di-nuclear systems as formed in the $^{28}$Si+$^{12}$C and $^{28}$Si+$^{28}$Si
reactions, by using the properties of emitted LCP.
Inclusive as well as exclusive data of the heavy fragments (A $\geq$ 6) and
their associated light charged particles (p, d, t, and $\alpha$-particles)
emitted in both reactions at $E_{lab}$($^{28}$Si) =112, 180 MeV have been measured
~\cite{bhattacharya99,rou200,bec200,rouph} using the {\bf ICARE}
 multidetector array ~\cite{bellot97}. The
LCP's emitted from FF fragments may provide the deformation properties of these
fragments. Moreover, the In-plane angular correlations data will be used to
extract the emitters temperatures. In this paper we will present the
exclusive data of LCP's in coincidence with ER. 

\vspace{-.3cm}

\section{Experimental Details}

\vspace{-.25cm}
\noindent
The two experiments were performed at the IReS Strasbourg VIVITRON Tandem
facility using 112 MeV and 180 MeV $^{28}$Si beams on $^{12}$C (160
$\mu$g/cm${^2}$ thick) and $^{28}$Si (230 $\mu$g/cm${^2}$ thick) targets
respectively~\cite{bhattacharya99,rou200,bec200,rouph}.
Both the heavy ions (A $\geq$ 6) and their associated LCP's (p, d, t, and
$\alpha$) were detected using the {\bf ICARE} charged particle multidetector
array~\cite{bellot97} which consists in nearly 40 telescopes in coincidence. The heavy
fragments, i.e. ER, quasi-elastic (QE), DI, and FF fragments, were detected in
10 telescopes, each consisting of an ionisation chamber (IC) followed by a 500
$\mu$m Si detector. The in-plane coincident LCP's were detected using either 3
three-elements telescopes (Si 40 $\mu$m, Si 300 $\mu$m, 2 cm CsI(Tl)) or 24
two-elements telescopes (Si 40 $\mu$m, 2 cm CsI(Tl)) and two double telescopes
(IC, Si 500 $\mu$m) located at the most backward angles. The IC's were filled
with isobutane and the pressures were kept at 30 Torrs and at 60 Torrs for
detecting heavy fragments and light fragments, respectively. The acceptance
of each telescope was defined by thick aluminium collimators.

\begin{figure}[htbp]
\epsfxsize=7.3cm
\centerline{\epsfbox{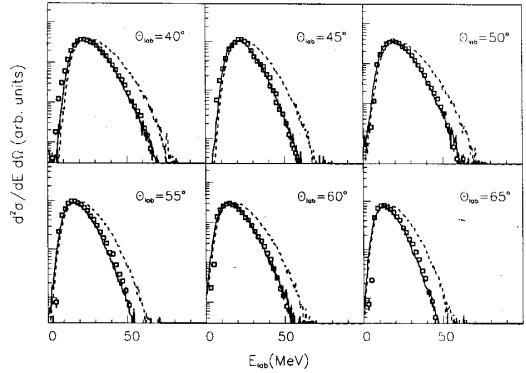}}
\caption{\it exclusive energy spectra of $\alpha$-particles
emitted, at the indicated laboratory angles, in the 180 MeV $^{28}$Si+$^{28}$Si
reaction. The solid and dashed lines are statistical-model calculations
discussed in the text.} 
\label{figure2}
\end{figure}

\vspace{-1.0cm}

\section{Experimental Results}

\vspace{-.2cm}

\subsection {Exclusive Energy spectra}

\vspace{-.4cm}
\noindent
The LCP spectra obtained in  both  of the reactions have Maxwellian shapes with
an exponential fall-off at high energy (their shape and high-energy slopes are
essentially independent of c.m. angle) which reflects a relatively low
temperature of the decaying nucleus.

\noindent
This is the signature of a statistical de-excitation process arising from a thermalized compound nucleus like source.
The energy spectra of the coincident $\alpha$-particles, for the reactions $^{28}$Si+$^{28}$Si and $^{28}$Si+$^{12}$C at
E$_{lab}$=180MeV, have been displayed in Figs.~1 and 2 respectively~\cite{bec200}.\\

\begin{figure}[htbp]

\epsfxsize=7.5cm
\centerline{\epsfbox{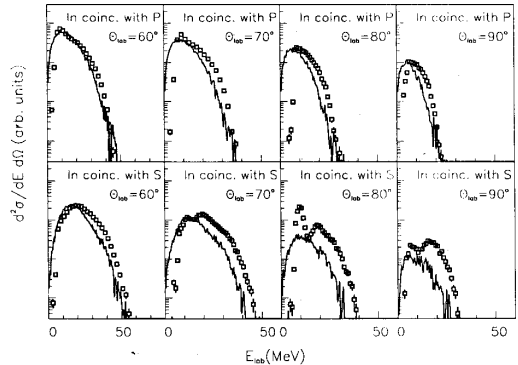}}
\caption{\it Exclusive energy spectra of $\alpha$-particles
emitted in coincidence with S and P ER's detected at -10$^\circ$, at the
indicated laboratory angles, in the 180 MeV $^{28}$Si+$^{12}$C reaction. The
solid lines are statistical-model calculations discussed in the text.} 
\label{comp2}
\end{figure}

\vspace{-.95cm}

\subsection{Analysis and Discussions}

The analysis of the data has been performed using {\sc CACARIZO}~\cite{viesti88}, the
Monte Carlo version of the statistical-model code {\sc CASCADE}~\cite{puhlhofer77}
. The informations on the main
ingredients of the statistical description, {\it i.e.,} the nuclear level
densities and the barrier transmission probabilities, are usually obtained from
the study of the evaporated light particle spectra. In recent years, it has
been observed that the standard statistical model could not predict
satisfactorily, the shape of the evaporated $\alpha$-particle energy
spectra~\cite{viesti88,govil87,larana88,huizenga89,govil98} and the measured
average energies of the $\alpha$-particles were much lower than the
corresponding theoretical predictions. Several attempts have been made in the
past few years to explain this anomaly either by changing the emission barrier
or using a spin-dependent level density~\cite{viesti88,govil87,huizenga89,govil98}.
 In hot rotating nuclei formed in heavy-ion reactions,
the energy level density at higher angular momentum is spin dependent.
 The level density, $\rho(E,J)$, for a given angular momentum $J$ and energy $E$ is
given by the well known Fermi gas expression : 

\begin{equation}
\hspace{-1.7cm} \rho(E,J) = {\frac{(2J+1)}{12}}a^{1/2}
           ({\frac{ \hbar^2}{2 {\cal J}_{eff}}}) ^{3/2}
           {\frac{1}{(E-\Delta-t-E_J)^2} }exp(2[a(E-\Delta-t-E_J)]^{1/2})
\label{lev}
\end{equation}

\noindent
where $a$ the level density parameter is taken equal to A/8, t is the
``nuclear" temperature, and $\Delta$ is the pairing correction, E$_J$ = $\frac{
\hbar^2}{2 {\cal J}_{eff}}$J(J+1) is the rotational energy, ${\cal J}_{eff}=
{\cal J}_0 \times (1+\delta_1J^2+\delta_2J^4)$ is the effective moment of
inertia,  ${\cal J}_0$ is the rigid body moment of inertia, and $\delta_1$ and
$\delta_2$ are deformation parameters~\cite{puhlhofer77}. 

\noindent
By changing the deformability parameters $\delta_1$ and $\delta_2$ one can
simulate the spin-dependent level density~\cite{viesti88,govil87,huizenga89}.
The CACARIZO calculations have been performed using two sets of input
parameters: one with a standard set consistent with the deformation of the
finite range liquid drop model (FRLDM)~\cite{sierk}, and another with a
spin-dependent moment of inertia and non-zero values for the deformability
parameters.

\noindent
As observed at E$_{lab}$= 112 MeV data ~\cite{bhattacharya99}, the energy spectra of the $\alpha$ spectra is
well reproduced after including the deformation. The dashed lines
in Fig.~1 show the 
predictions of {\sc CACARIZO} using the standard FRLDM deformation parameter
 set. It is clear that the
average energies of the measured $\alpha$ energy spectra are lower than those
predicted by the standard statistical-model calculations. The solid lines show
the predictions of {\sc CACARIZO} using the paramater set with non- zero
values $\delta_1$ and $\delta_2$.

\noindent
 The exclusive energy spectra
of $\alpha$-particle measured in coincidence with individual S and P ER's,
which are shown in Fig.~2 for the reaction $^{28}$Si+$^{12}$C at
E$_{lab}$ = 180 MeV, are quite interesting. The
spectra associated with S are completely different from those associated with
P. The latter are reasonably well reproduced by the CACARIZO curves whereas the
model could not predict the shape of the spectra obtained in coincidence with
S. An additional non-statistical components appear to be significant in this
case.  The similar results
have been observed at the lower bombarding energy E$_{lab}$ = 112 MeV
~\cite{bhattacharya99,rou200}.
These non-statistical components may come from the decay
of unbound $^{8}$Be (in g.s. and the two first excited states) produced in
the binary reaction $^{40}$Ca $\rightarrow$ $^{32}$S+$^{8}$Be of a massive transfer
 type~\cite{arena}.\\

\end{document}